\documentclass[preprint,12pt]{elsarticle}%
\usepackage{amssymb}
\usepackage{amsfonts}
\usepackage{amsmath}%
\usepackage{graphicx}
\providecommand{\U}[1]{\protect\rule{.1in}{.1in}}
\graphicspath{{E:/dottorato/Coulomb_damper/Testo/Paper/Immagini/}}
\journal{journal}

\begin{document}
\bigskip%
\begin{frontmatter}%


%

\title{Adhesion of surfaces with wavy roughness and a shallow depression }%

%

\author{A. Papangelo(1,2), M. Ciavarella(1) }%
%

\address
{(1) Politecnico di BARI. Center of Excellence in Computational Mechanics. Viale Gentile 182, 70126 Bari. Mciava@poliba.it
(2) Department of Mechanical Engineering, Hamburg University of Technology, Am Schwarzenberg-Campus 1, 21073 Hamburg, Germany}%
%

\begin{abstract}%

Recently, a\ simple and elegant "dimple" model was introduced by McMeeking
\textit{et al.} (Adv\ Eng Mat 12(5), 389-397, 2010) to show a mechanism for a
bistable adhesive system involving a surface with a shallow depression. The
system shows, at least for intermediate levels of stickiness, that external
pressure can switch the system into a "strong adhesive" regime of full
contact, or into weak adhesion and complete pull-off, similarly to the contact
of surfaces with a single scale of periodical waviness. We add to this model
the effect of roughness, in the simple form of axisymmetric single scale of
waviness, permitting a very detailed study, and we show that this induces a
resistance to jumping into full contact on one hand (limiting the "strong
adhesion" regime), and an enhancement of pull-off and of hysteresis starting
from the partial contact state on the other (enhancing the "weak adhesion"
regime). We show the system depends only on two dimensionless parameters,
depending on the ratio of work of adhesion to the energy to flatten the dimple
or waviness, respectively. The system becomes pressure-sensitive also in the
intermediate states, as it is observed in real adhesive rough systems. The
model obviously is specular to the Guduru model of rough spheres (Guduru,
JMPS, 55, 473--488, 2007), with which it shares the limitations of the
analysis assuming a connected contact\ (crack) area, and serves also the
purpose of showing the effect of a depression into an otherwise periodic rough
contact, towards the understanding of adhesion with multiple scales of roughness.%

\end{abstract}%
%

%

\end{frontmatter}%



\section{\bigskip Introduction}

Adhesion in the presence of roughness is usually destroyed very easily, as it
was proved by Fuller and Tabor (1975), even in low modulus materials like
smooth rubber lenses against roughened surfaces. Therefore, it is still
surprising that some insects use adhesion for their locomotion, by using a
series of mechanisms, the study of which has generated a very important area
of research in the last decades. The mechanisms include splitting the contact
into many spots and optimizing the shape and size of each contact (Hui
\textit{et al.} 2004, Kamperman \textit{et al.}, 2010, Gao \& Yao 2004).
However, even the best of the mechanisms, is unlikely to work with all
possible rough surfaces, showing wavelengths and amplitude over different
length scales (Huber \textit{et al.}, 2007, Pugno \&\ Lepore, 2008), showing a
truly efficient system for multiscale arbitrary roughness is extremely
difficult to achieve. The understanding about when adhesion can be very strong
or very weak depending on features of roughness, pre-load, and system
architecture is so far very remote from being complete.

At the opposite end of the classical finding of Fuller \&\ Tabor (1975),
Johnson (1995) demonstrated a mechanism for which roughness in the form of a
sinusoidal wave, has a \textit{minimal effect }after a sufficiently high
pressure has been applied, because the contact naturally jumps into a state of
full contact.\ Indeed, this can happen even \textit{spontaneously} (at zero
external load) for sufficiently high work of adhesion. After this state has
been reached, virtually the theoretical strength of the material is found, and
one has to postulate either a tiny flaw at the interface, or air entrapment to
escape this limit which is far from common experience. Therefore, the role of
roughness can be pressure-sensitive.

Guduru (2007) found a mechanism of enhancement which has some connection with
Johnson's model, in that he imagined a sphere with roughness in the form of
axisymmetric waviness, and solved the problem assuming that roughness was
effectively flattened during the deformation, so the contact was a simply
connected area. This results in very large oscillations in the normal load as
a function of indentation, which were also observed experimentally in Guduru
\&\ Bull (2007), and gave rise to both large dissipation because of multiple
jumps from unstable to stable branches, and to an enhancement of pull-off.
Kesari \&\ Lew (2010) further discussed how these unstable jumps could in the
limit of small wavelength roughness define a continuous curve removing the
oscillations of the original solution: Ciavarella (2016a) further remarked
that the Kesari \&\ Lew (2010) asymptotic expansion corresponds to splitting
the classical JKR theory solution for spheres given in (Johnson \textit{et
al.}, 1971) into two branches, loading and unloading ones, which correspond to
an decreased and increased values of work of adhesion, respectively, uniquely
dependent on the Johnson (1995) parameter for sinusoidal waviness contact.
Ciavarella (2016b) further used the Kesari \&\ Lew (2010) asymptotic expansion
for multiscale roughness in the form of a Weierstrass function, showing that
the enhancement could be extremely large in this case, although the
assumptions of the simply connected area solution become increasingly stretched.

Recently, McMeeking \textit{et al.} (2010) have proposed a very simple model
where two surfaces are gently brought into contact, one of which having a
single small depression. This is in a sense a simplification of the single
scale of waviness of Johnson's (1995) model, with the additional significant
advantage that in full 3D situations, sinusoidal roughness leads in
intermediate regimes to a very complex problem with non-circular contact
areas, whereas the dimple model preserves axisymmetry and permits a very
simple solution, particularly with the shape chosen by McMeeking \textit{et
al.} The model however preserves all of the features of the periodic waviness
problem, in that there is a possibility of jump-into contact at some level of
compression (or it can be spontaneous for sufficient level of adhesion), and
that there is an unstable pull-off at some value of external tension. This
pull-off is no longer occurring on the crests, and therefore is not the known
value for spheres given by JKR theory (Johnson \textit{et al.}, 1971), but
depends on the shape of the depression.

As Johnson (1995) remarked, the single scale waviness model shows an extreme
behaviour, which is likely to be affected by deviation due also to the
presence of finer-scale roughness. Guduru's model already answers some
questions about the case of two scales of waviness, for contact near the
crest, but the spherical geometry doesn't admit a "full contact" limit, and
therefore it doesn't address the problem towards this regime. Instead, the
simple dimple model of McMeeking \textit{et al.} (2010) is ideal to a
quantitative assessment of the full problem, by adding axisymmetric roughness.
In other words, despite idealized, this geometry can be a model for two scales
of roughness both in what happens in a single scale of waviness when
encountering a depression in a surface, or specularly, for the depression
itself having roughness. The advantage of the model is that we can study it in
great details, especially in the asymptotic regime in which the wavelength of
the roughness is much smaller than the non-contact area. We hope to elucidate
qualitatively some features of the general behaviour of adhesive surfaces.

\section{Formulation}

We consider as smooth geometry that of a surface with a shallow depression, in
the form of a dimple of amplitude $\delta_{0}$ and radius $b$ (McMeeking
\textit{et al.} 2010), defined as%

\begin{align}
\delta & =\frac{2}{\pi}\delta_{0}E\left(  \frac{r}{b}\right)  ,\text{
\ \ \ \ \ \ \ }\frac{r}{b}\leq1\\
\delta & =\frac{2}{\pi}\delta_{0}\frac{r}{b}\left[  E\left(  \frac{r}%
{b}\right)  -\left[  1-\left(  \frac{b}{r}\right)  ^{2}\right]  K(\frac{b}%
{r})\right]  ,\text{ \ \ \ \ \ \ \ }\frac{r}{b}>1
\end{align}
where $K,E$ are complete elliptic integrals of first and second kind. To the
smooth geometry, we add an axisymmetric sinusoidal roughness of amplitude and
wavelength $g,\lambda$.

The geometry is clarified in Fig.1.

\begin{center}
$%
\begin{array}
[c]{cc}%
{\includegraphics[
height=2.565in,
width=4.0274in
]%
{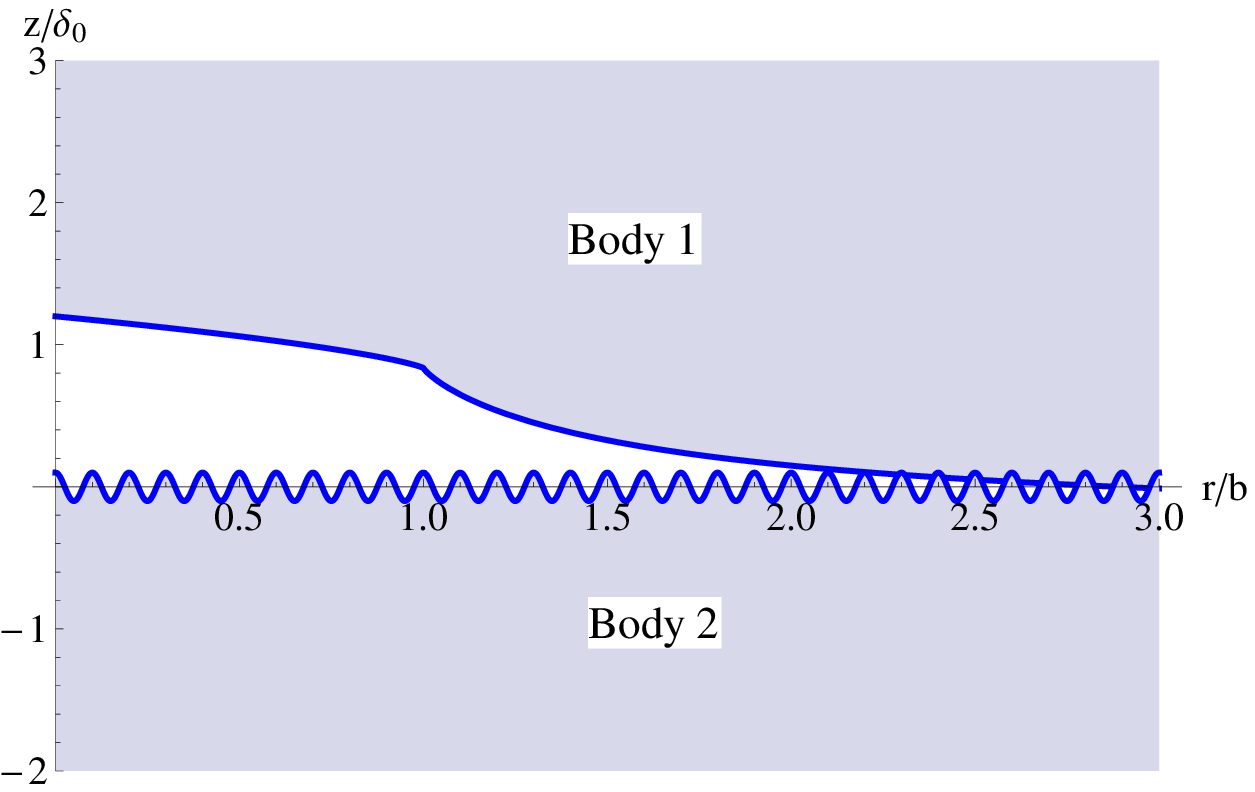}%
}
&
\end{array}
$

Fig.1 Geometry of the system. A flat surface having a shallow depression in
the form of a "dimple" -- against a surface with wavy roughness (roughness
could be added to either of the surfaces).
\end{center}

The system will tend to adhere from the remote points towards the center, and
therefore, assuming the contact is continuous (which requires roughness is not
too large, see Discussion), the contact correspond to a classical penny-shaped
crack. Using superposition principles, we can solve for the penny-shaped crack
under internal tension, given by the state of stress induced in the full
adhered state. The latter is defined by the combination of 3 components:

\begin{itemize}
\item (i) the localized tensile stress $T$ in $r<b$, i.e. inside the dimple
\begin{equation}
T=\frac{E^{\ast}\delta_{0}}{2b},\qquad r<b
\end{equation}

\item (ii) a remote tension $\sigma_{A}$ in the entire plane, and

\item (iii) a sinusoidal variation
\begin{equation}
\sigma_{rough}=p^{\ast}\cos\frac{2\pi r}{\lambda}\label{plane-strain}%
\end{equation}
where $p^{\ast}=\pi E^{\ast}g/\lambda$. Notice this is a plane strain
approximation, which is proven to be extremely good for our scopes in Appendix.
\end{itemize}

Using the standard results for axisymmetric cracks (see eg. Maugis, 2000)
under internal pressure $p\left(  r\right)  $, for a crack of radius $c$, we
derive the stress intensity factor as (Maugis, 2000, 3.117)
\begin{equation}
K_{I}=\frac{2}{\sqrt{\pi c}}\int_{0}^{c}\frac{sp\left(  s\right)  ds}%
{\sqrt{c^{2}-s^{2}}}%
\end{equation}

The first two components (i-ii) give
\begin{align}
K_{I}  & =\left(  \sigma_{A}+T\right)  \frac{2\sqrt{c}}{\sqrt{\pi}}\text{, for
}c<b\label{KI1}\\
& =\frac{2}{\sqrt{\pi c}}\left[  \sigma_{A}c+T\left(  c-\sqrt{c^{2}-b^{2}%
}\right)  \right]  \text{, for }c>b\label{KI2}%
\end{align}

The third component (iii) gives
\begin{equation}
K_{I}=\frac{2p^{\ast}}{\sqrt{\pi c}}\int_{0}^{c}\frac{s\cos\frac{2\pi
s}{\lambda}ds}{\sqrt{c^{2}-s^{2}}}=p^{\ast}\sqrt{\pi c}H_{-1}\left(
\frac{2\pi c}{\lambda}\right) \label{KI3}%
\end{equation}
where $H_{-1}$ is the Struve function of order -1.

Under the assumptions of Linear Elastic Fracture Mechanics (often in adhesion,
called "JKR" regime, from Johnson \textit{et al.}, 1971), equilibrium is
obtained equating the total $K_{I}$ to the $K_{Ic}$ toughness of the material
pair, or equivalently the energy release rate to the work of adhesion $G=w$
where $2E^{\ast}w=K_{Ic}^{2}$. Hence we obtain from (\ref{KI1}, \ref{KI2},
\ref{KI3})
\begin{align}
K_{Ic}  & =\left(  \sigma_{A}+T\right)  \frac{2\sqrt{c}}{\sqrt{\pi}}+p^{\ast
}\sqrt{\pi c}H_{-1}\left(  \frac{2\pi c}{\lambda}\right)  \text{, for }c<b\\
K_{Ic}  & =\frac{2}{\sqrt{\pi c}}\left[  \sigma_{A}c+T\left(  c-\sqrt
{c^{2}-b^{2}}\right)  \right]  +p^{\ast}\sqrt{\pi c}H_{-1}\left(  \frac{2\pi
c}{\lambda}\right)  \text{, for }c>b
\end{align}
i.e.
\begin{align}
\frac{\sigma_{A}}{T}  & =-1-\frac{\pi}{2}\frac{p^{\ast}}{T}H_{-1}\left(
\frac{2\pi c}{\lambda}\right)  +\frac{\pi}{2}\frac{K_{Ic}}{T\sqrt{\pi c}%
}\text{, for }c<b\\
\frac{\sigma_{A}}{T}  & =-\left(  1-\sqrt{1-\left(  \frac{b}{c}\right)  ^{2}%
}\right)  -\frac{\pi}{2}\frac{p^{\ast}}{T}H_{-1}\left(  \frac{2\pi c}{\lambda
}\right)  +\frac{\pi}{2}\frac{K_{Ic}}{T\sqrt{\pi c}}\text{, for }c>b
\end{align}

One natural dimensionless parameter is then%
\begin{equation}
\alpha_{d}=\frac{K_{Ic}\sqrt{\pi}}{2T\sqrt{b}}=\frac{\pi K_{Ic}}{2T\sqrt{\pi
c}}\frac{\sqrt{c}}{\sqrt{b}}\label{alfad}%
\end{equation}
which, by analogy with the sinusoidal case of Johnson (1995), we can define
the Johnson parameter for the smooth dimple, proportional to the ratio between
the work of adhesion, and the elastic energy to flatten the smooth dimple.
Hence, we can restate the LEFM curves normalizing the stresses by $T$ as
$\widehat{\sigma}_{A}=\frac{\sigma_{A}}{T}$ and $\widehat{p}^{\ast}=$
$\frac{p^{\ast}}{T}$, and all length scales by $b$ as $\widehat{c}=c/b$ and
$\widehat{\lambda}=\lambda/b$, obtaining
\begin{align}
\widehat{\sigma}_{A}  & =-1-\frac{\pi}{2}\widehat{p}^{\ast}H_{-1}\left(
\frac{2\pi\widehat{c}}{\widehat{\lambda}}\right)  +\frac{\alpha_{d}}%
{\sqrt{\widehat{c}}}\text{, for }\widehat{c}<1\\
\widehat{\sigma}_{A}  & =-1+\sqrt{1-\frac{1}{\widehat{c}^{2}}}-\frac{\pi}%
{2}\widehat{p}^{\ast}H_{-1}\left(  \frac{2\pi\widehat{c}}{\widehat{\lambda}%
}\right)  +\frac{\alpha_{d}}{\sqrt{\widehat{c}}}\text{, for }\widehat{c}>1
\end{align}

Using the expansion of $H_{-1}\left(  \frac{2\pi\widehat{c}}{\widehat{\lambda
}}\right)  $ for $\widehat{\lambda}\rightarrow0$, similarly to what done by
Kesari \&\ Lew (2011) for the problem of a rough sphere, we obtain the first
term as\footnote{This was obtained with some simple manipulations of the
Mathematica command "Series".}
\[
H_{-1}\left(  \frac{2\pi\widehat{c}}{\widehat{\lambda}}\right)  \simeq\frac
{1}{\pi}\sqrt{\frac{\widehat{\lambda}}{\widehat{c}}}\sin\left(  \frac{\pi}%
{4}+\frac{2\pi\widehat{c}}{\widehat{\lambda}}\right)
\]
and therefore, this function will have minima (and maxima) which can be
joined, obtaining%
\begin{align}
\widehat{\sigma}_{A}^{\pm}  & =-1\pm\frac{\widehat{p}^{\ast}}{2}\sqrt
{\frac{\widehat{\lambda}}{\widehat{c}}}+\frac{\alpha_{d}}{\sqrt{\widehat{c}}%
}\text{, for }\widehat{c}<1\\
\widehat{\sigma}_{A}^{\pm}  & =-1+\sqrt{1-\frac{1}{\widehat{c}^{2}}}\pm
\frac{\widehat{p}^{\ast}}{2}\sqrt{\frac{\widehat{\lambda}}{\widehat{c}}}%
+\frac{\alpha_{d}}{\sqrt{\widehat{c}}}\text{, for }\widehat{c}>1
\end{align}

Exactly as it happens for the rough sphere (see Ciavarella, 2016a, 2016b), we
can group the terms corresponding to the smooth profile and the roughness
induced first order increase, and accordingly define an increased/decreased
work of adhesion for unloading/loading respectively. For the rough sphere,
this results in JKR equations with different work of adhesion on loading and
unloading. Here, this results in%

\begin{align}
\widehat{\sigma}_{A}^{\pm}  & =-1+\frac{1}{\sqrt{\widehat{c}}}\left(
\alpha_{d}\pm\frac{\widehat{p}^{\ast}}{2}\sqrt{\widehat{\lambda}}\right)
\text{, for }\widehat{c}<1\label{env1}\\
\widehat{\sigma}_{A}^{\pm}  & =-1+\sqrt{1-\frac{1}{\widehat{c}^{2}}}+\frac
{1}{\sqrt{\widehat{c}}}\left(  \alpha_{d}\pm\frac{\widehat{p}^{\ast}}{2}%
\sqrt{\widehat{\lambda}}\right)  \text{, for }\widehat{c}>1\label{env2}%
\end{align}

Hence, effectively the curves are identical to those of the "smooth" dimple,
where one substitutes $\alpha_{d}$ with
\begin{align}
\alpha_{d,rough}^{loading}  & =\left(  \alpha_{d}-\frac{\widehat{p}^{\ast}}%
{2}\sqrt{\widehat{\lambda}}\right)  =\alpha_{d}\left(  1-\frac{1}{\sqrt{\pi
}\alpha}\right) \\
\alpha_{d,rough}^{unloading}  & =\left(  \alpha_{d}+\frac{\widehat{p}^{\ast}%
}{2}\sqrt{\widehat{\lambda}}\right)  =\alpha_{d}\left(  1+\frac{1}{\sqrt{\pi
}\alpha}\right)
\end{align}
where we have recognized $\alpha=\frac{2\alpha_{d}}{\widehat{p}^{\ast}%
\sqrt{\pi\widehat{\lambda}}}$ with $\alpha$ being exactly the parameter
introduced by Johnson (1995) for the roughness of a single sinusoid of
wavelength and amplitude $\lambda,g$, defined as
\begin{equation}
\alpha^{2}=\frac{2}{\pi^{2}}\frac{w\lambda}{E^{\ast}g^{2}}\label{alfa-KLJ}%
\end{equation}

Notice that while $\alpha_{d,rough}^{unloading}$ is always greater than the
original $\alpha_{d}$, $\alpha_{d,rough}^{loading}$ is always smaller, and
indeed it can become negative (for $\alpha<1/\sqrt{\pi}=0.57$). This simply
means that the curve for loading will not correspond to any physical curve for
a smooth dimple, and in fact it is due to the fact that the pressure to reach
full contact becomes even higher than that of the smooth dimple without adhesion.

\section{Results}

To elucidate the results, let us start by showing the curves for the "smooth"
dimple, in Fig.2. In particular, Fig.2a shows various values of dimple
parameter $\alpha_{d}$ , from zero (adhesionless contact), which requires a
compression $\widehat{\sigma}_{A}=-1$ to obtain full contact, to values higher
than 1, which show no equilibrium points at zero external load -- implying
spontaneous jump into full contact. More precisely, considering the case
$\alpha_{d}=0.75$ plotted in Fig.2b, and starting from very large $\widehat
{c},$ equilibrium is achieved at point A.\ After that, if we load in
compression, we move towards negative $\widehat{\sigma}_{A}$ and jump into
full contact at point B, as the branch with negative slope is unstable. From
this full contact condition, only very high values of tension will detach the
interface, of the order of theoretical strength. If, instead, from point B we
start unloading, we enlarge the crack size until another instability is
reached, that of pull-off at point C.

\begin{center}%
\begin{tabular}
[c]{ll}%
{\includegraphics[
height=2.5218in,
width=4.0274in
]%
{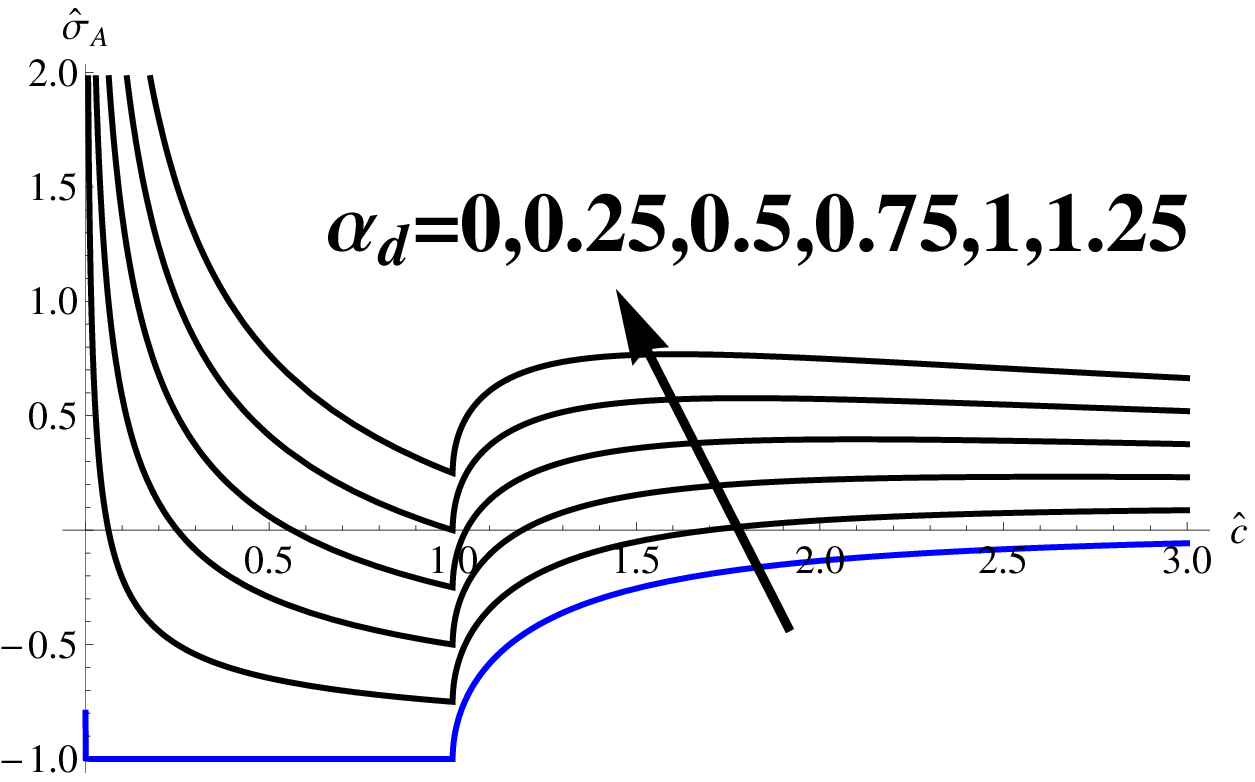}%
}
& (a)\\%
{\includegraphics[
height=2.4881in,
width=4.0274in
]%
{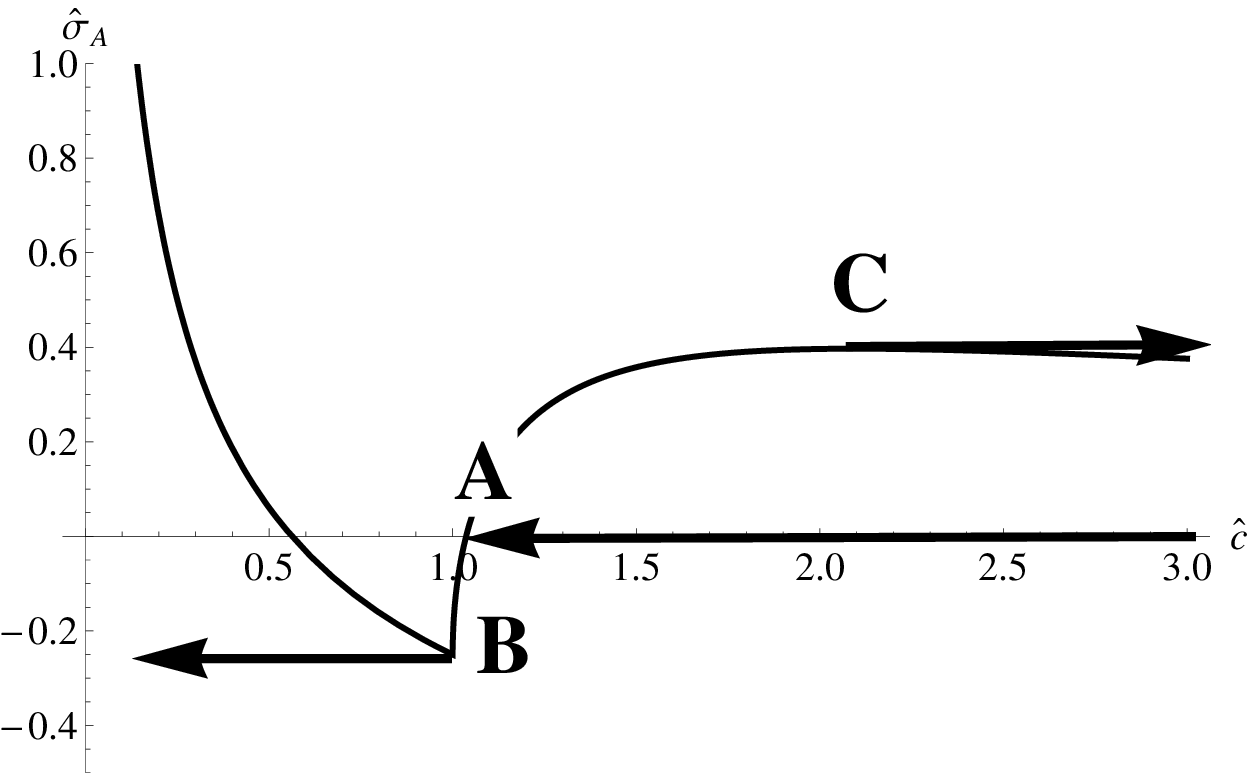}%
}
& (b)
\end{tabular}

Fig.2 Example results for the smooth dimple, (a) with $\alpha_{d}%
=0,0.25,0.5,0.75$. and (b) for the case $\alpha_{d}=0.75$ with various
loading/unloading paths%

\begin{tabular}
[c]{ll}%
{\includegraphics[
height=2.4881in,
width=4.0274in
]%
{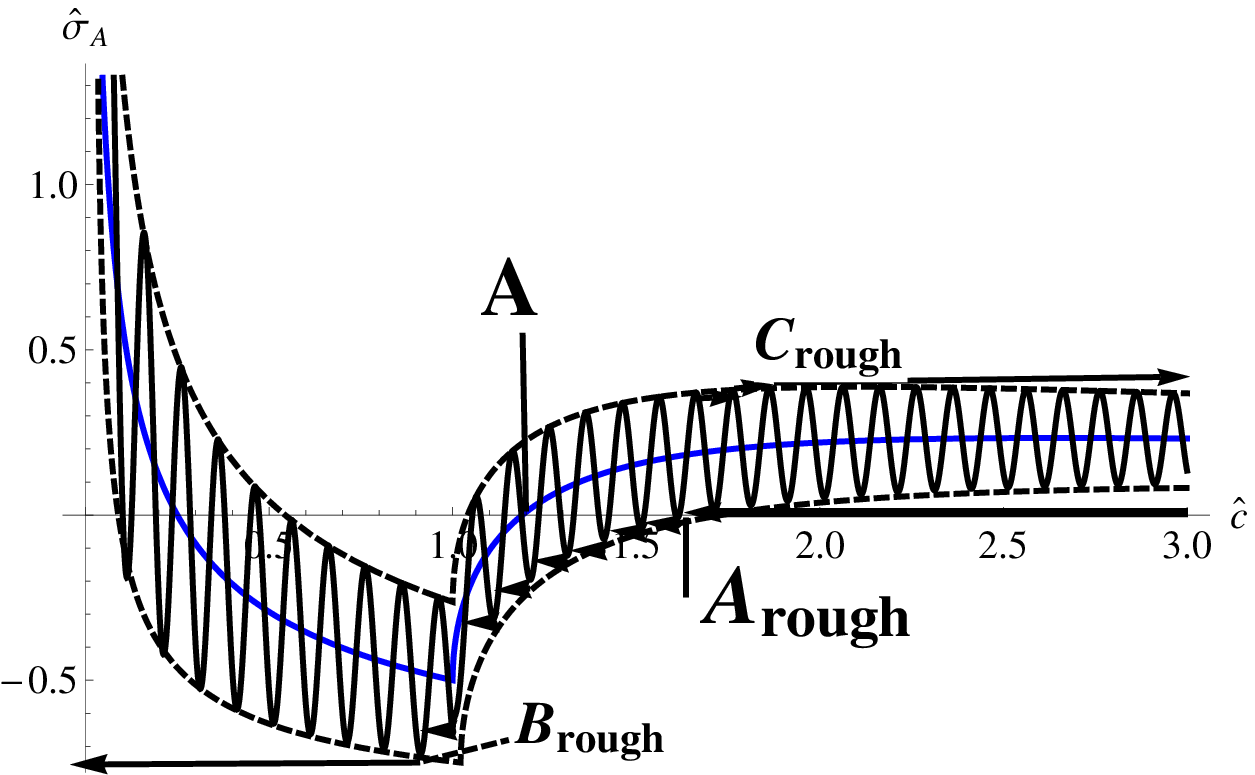}%
}
& (a)\\%
{\includegraphics[
height=2.5313in,
width=4.0274in
]%
{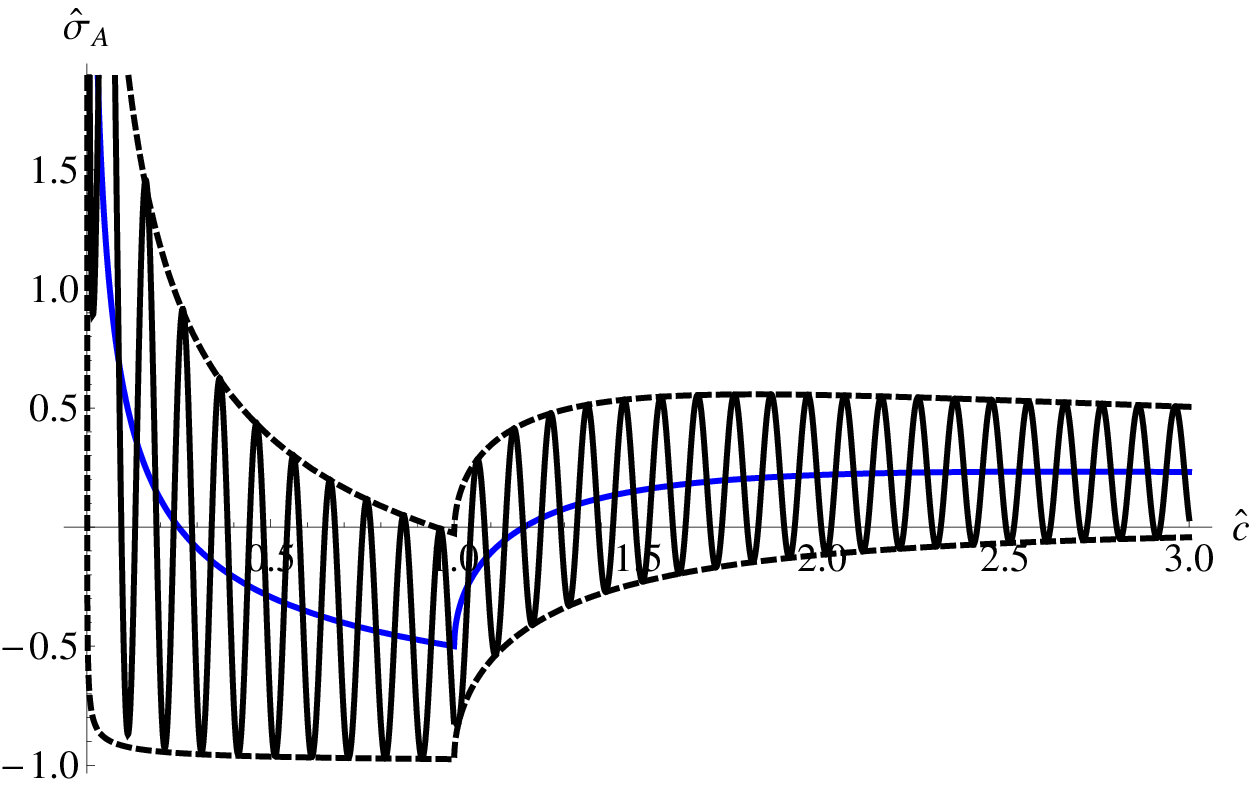}%
}
& (b)\\%
{\includegraphics[
height=2.5547in,
width=4.0274in
]%
{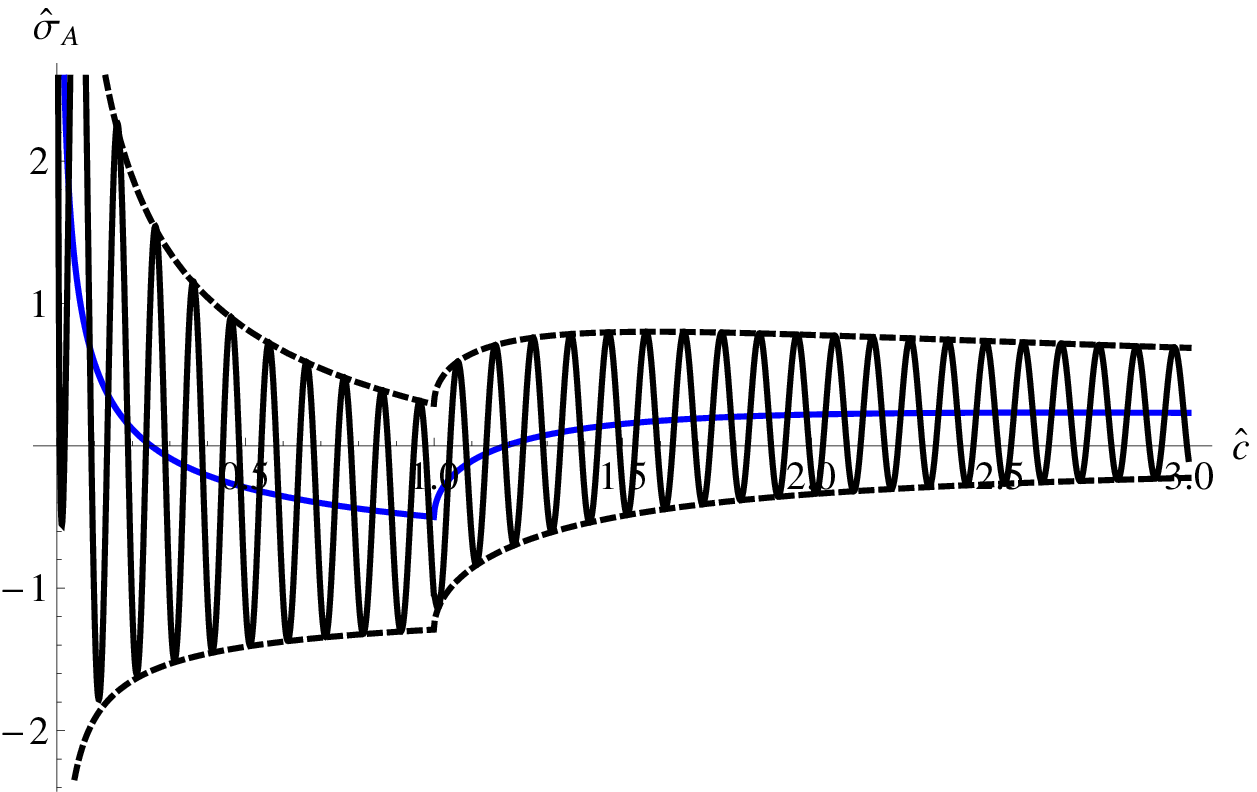}%
}
& (c)
\end{tabular}

Fig.3 Example results for the rough dimple $\alpha_{d}=0.5,\widehat{\lambda
}=0.1$. (a) $\widehat{p}^{\ast}=1.5,$ (and hence $\alpha=1.19$, $\alpha
_{d,rough}^{loading}=0.26,\alpha_{d,rough}^{unloading}=0.737$); (b)
$\widehat{p}^{\ast}=3,\ $ (and hence $\alpha=0.595$, $\alpha_{d,rough}%
^{loading}=0.026,\alpha_{d,rough}^{unloading}=0.97$); (c) $\widehat{p}^{\ast
}=5,$ (and hence $\alpha=0.357$, $\alpha_{d,rough}^{loading}=-0.29,\alpha
_{d,rough}^{unloading}=1.29$)
\end{center}

\bigskip Looking at the rough system, we show some examples in Fig.3, where we
plot with solid blue line the smooth system, with dashed lines the "envelope"
solutions (\ref{env1}, \ref{env2}). Starting with Fig.3a, we have a system
with $\alpha_{d}=0.5,\widehat{p}^{\ast}=1.5,\widehat{\lambda}=0.1 $ (and hence
$\alpha=1.19$) where we have evidenced with arrows the jumps of the system
when it encounters an unstable branch of the solution. Clearly, the
equilibrium crack size at zero external load, point A, has now moved to
A$_{rough}$ which corresponds to a larger size. Also, the jump into full
contact B$_{rough}$, occurs for a much higher compressive stress than in the
smooth case, and for a crack size slightly smaller. In unloading from point
A$_{rough}$, it is clear that pull-off will occur quite soon in C$_{rough}$,
although for a much larger tensile stress than in the smooth case.\ The
behaviour of the system is however generally well represented by the two
envelope curves, corresponding to $\alpha_{d,rough}^{loading},\alpha
_{d,rough}^{unloading}$ respectively for loading and unloading, on which we
will build some results in the next paragraph, except for the details of what
happens near zero crack size.

The cases in Fig.3b,c are for $\widehat{p}^{\ast}=3,5$ and show an increasing
effect of roughness. We have that the equilibrium point A can correspond to
crack sizes larger than the point of maximum pull-off, so that the effective
pull-off is smaller than the highest point, although still larger than the
smooth case. We shall return on these important elements when we will compute
the pull-off value as a function of the system parameters (loading and
geometrical ones). Also, for the case (c) we have that the jump into full
contact, due to increased roughness, moves increasingly towards very small
crack sizes, and much higher values of compression, higher than in the smooth
case. In this case, also the envelope solution breaks down, since the crack
size becomes comparable to the wavelength of the sinusoidal roughness, and
hence while the envelope solution suggests full contact only at infinite
compression and zero crack, the actual system will show a true jump into
contact depending on the full solution.

Naturally, this behaviour also results in an increased hysteresis, since when
we follow some loading/unloading cycles (particularly from high pressures even
if not enough to jump into full contact), with all the unstable jumps which
dissipate energy in the form of elastic waves emitted in the surface, and this
can result in an higher dissipation than in the smooth dimple case. To compute
this with precision, one should integrate load-displacement curves.

\subsection{Using envelope curves alone}

If we make the approximation that the wavelength is very small with respect to
the dimple size, the envelope curves joining all maxima and minima of the true
solution are a good approximation, and permit a very simple analysis of the
behaviour: the effect of roughness corresponds to split the loading and
unloading curves into two curves for smooth dimples at different $\alpha_{d}
$. For example, Fig.4 shows the envelope loading and unloading curves for a
case with $\alpha_{d}=1$ which, in the smooth case, is unstable whereas in the
rough case, for $\alpha=0.5,1$ as in the figure, has an equilibrium point:
this corresponds to a crack size $\widehat{c}$ which becomes very large for
low $\alpha$, and at the same time the unloading curve becomes very high
(corresponding to a value which would not be of interest for a smooth dimple,
since it would be in that case all unstable). In other words, the system shows
that full contact becomes remote, and viceversa, one has a very
pressure-sensitive behaviour, as the highest pull-off becomes achievable only
for large enough preliminary pre-load.

\begin{center}%
{\includegraphics[
height=2.5547in,
width=4.0274in
]%
{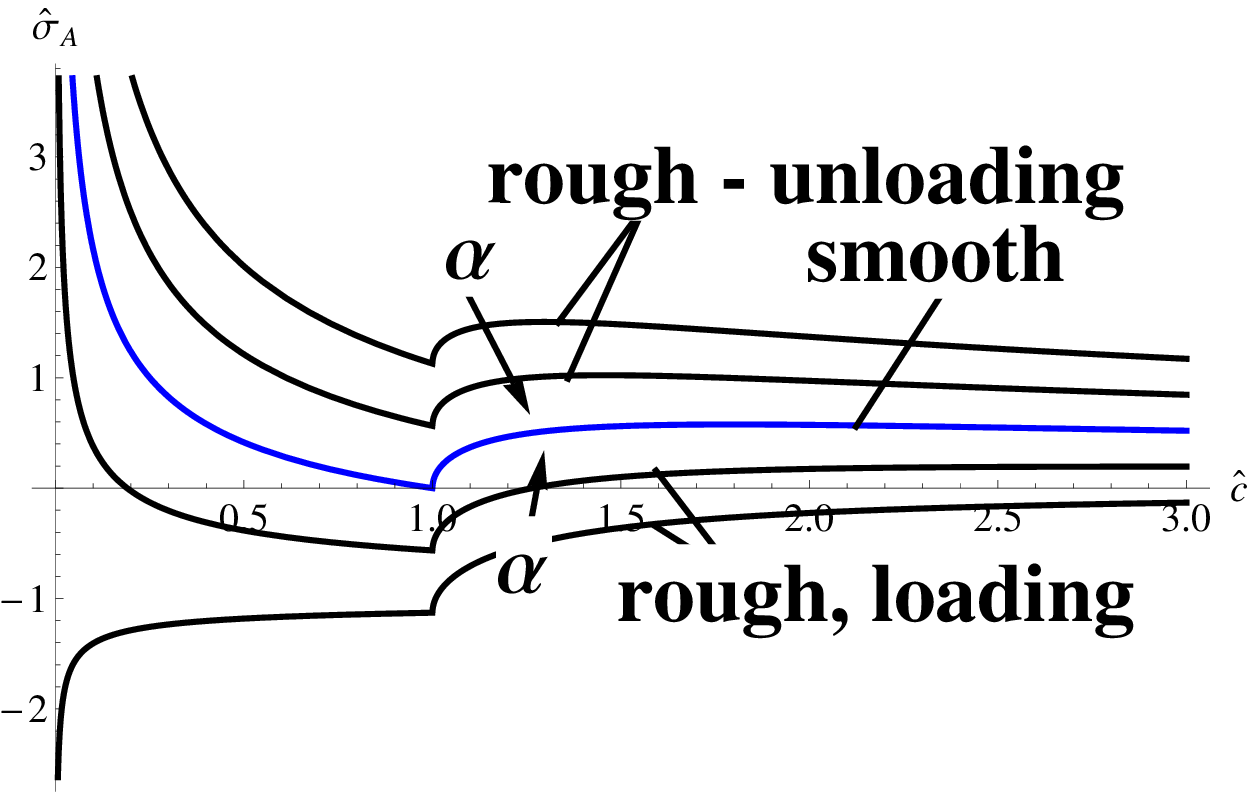}%
}

Fig.4. Equilibrium curves for smooth dimple with $\alpha_{d}=1$ (blue solid
line), together with envelope curves for rough dimple with $\alpha=0.5,1$ on
loading and unloading.
\end{center}

An aspect which is different from the Guduru-Kesari problem is that we have
shown how the effect of roughness tends to put some barrier to the jump into
full contact. We can assume that the condition of jump-into-full-contact
remains $\widehat{c}=1$: strictly this requires that $\alpha>\frac{1}%
{\sqrt{\pi}}=\allowbreak0.56$, but since after that the system is not governed
by the envelope solution, we shall make this approximation and therefore
estimate%
\begin{equation}
\widehat{\sigma}_{A}^{FC}\simeq-1+\alpha_{d,rough}^{loading}=-1+\alpha
_{d}\left(  1-\frac{1}{\sqrt{\pi}\alpha}\right)
\end{equation}
This is shown in Fig.5, where it is clear that we need to push as hard as in
the adhesionless condition, if $\alpha=\allowbreak0.56$; on the other hand,
that a partial contact state is found for a much wider range of conditions
than in the smooth case, and namely for $\widehat{\sigma}_{A}^{FC}<0$ which
occurs for
\begin{equation}
\alpha_{d}<\frac{1}{1-\frac{1}{\sqrt{\pi}\alpha}}\text{ or }\alpha>\frac
{1}{\sqrt{\pi}}=\allowbreak0.56\label{condition-partial-contact}%
\end{equation}
instead of $\alpha_{d}<1$ of the smooth case.

\begin{center}%
\begin{tabular}
[c]{ll}%
{\includegraphics[
height=2.5313in,
width=4.0274in
]%
{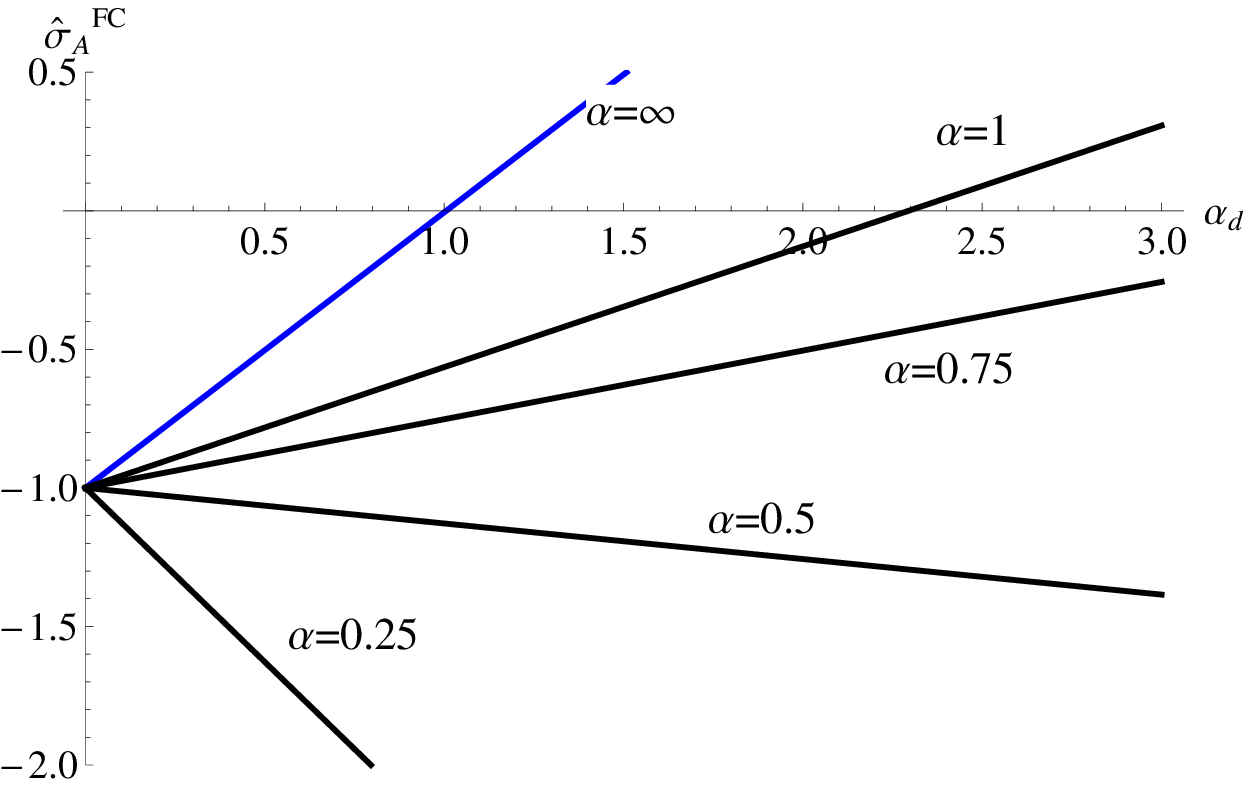}%
}
&
\end{tabular}

Fig.5. Tension to jump into full-contact $\widehat{\sigma}_{A}^{FC}$ as a
function of parameter $\alpha_{d}=1$ for various roughness parameter
$\alpha=0.25,0.5,0.75,1,\infty$ . The case $\alpha=\infty$ (blue solid line)
corresponds to the smooth dimple, whereas for the cases of small $\alpha<0.57$
we have taken an upper bound estimate considering the value reached at
$\widehat{c}=1$, as otherwise the actual minimum depends on the details of the
system, see also "Discussion".
\end{center}

\subsection{Upper bound to pull-off}

Moving to some consideration of pull-off, for the unloading curve from a
partial contact condition, we know the size of the contact for pull-off from
the solution of the smooth geometry (McMeeking \textit{et al.} 2010), which
however should now be computed for $\alpha_{d,rough}^{unloading}$. The full
expression is
\begin{equation}
\widehat{c}_{\max}=\left(  m+n\right)  ^{1/3}+\left(  m-n\right)  ^{1/3}%
\end{equation}
where $m=\frac{2}{\left(  \alpha_{d,rough}^{unloading}\right)  ^{2}}$, and
$n=\sqrt{m^{2}-\frac{1}{27}}$, but this result can be fitted very well with
\begin{equation}
\widehat{c}_{\max}\simeq1.82/\left(  \alpha_{d,rough}^{unloading}\right)
^{0.65}%
\end{equation}

A good expression for $\widehat{\sigma}_{A,smooth}^{PO}$ then turns out to be
($\alpha_{d}<1$ is the limit to have a partial contact state in the smooth
surface case)
\begin{equation}
\widehat{\sigma}_{A,smooth}^{PO}\simeq0.53\alpha_{d}^{1.3}<0.53
\end{equation}

Therefore, an upper bound to the pull-off for the rough case (where upper
bound means provided we have sufficient pre-load as to reach this stage) is
\begin{equation}
\widehat{\sigma}_{A,rough}^{PO\max}=0.53\left(  \alpha_{d,rough}%
^{unloading}\right)  ^{1.3}=0.53\left(  \alpha_{d}\left(  1+\frac{1}{\sqrt
{\pi}\alpha}\right)  \right)  ^{1.3}\label{sigma-pull-off-max}%
\end{equation}
and hence the enhancement is really defined by the ratio
\begin{equation}
\frac{\widehat{\sigma}_{A,rough}^{PO\max}}{\widehat{\sigma}_{A,smooth}^{PO}%
}=\left(  1+\frac{1}{\sqrt{\pi}\alpha}\right)  ^{1.3}%
\end{equation}
but this time the expression holds as long as we satisfy the condition for
partial contact (\ref{condition-partial-contact}). In Fig.6a, we plot for
representative cases the value of $\widehat{\sigma}_{A,rough}^{PO\max}$
together with $\widehat{\sigma}_{A,smooth}^{PO}$ (which corresponds to
$\alpha=\infty$ and is plotted with blue solid line). As it is clear, already
for $\alpha=1$ we have a significant increase (about 77\%), but this holds
only until we have the possibility of partial contact solution (this is
indicated with arrows going upwards indicating the regime of full contact and
"strong adhesion"). However, it is with $\alpha<0.57$ that the increase
becomes really substantial and in principle without limit (see the Discussion
paragraph). Supposing we can apply the solution for $\alpha=0.25$, this
corresponds to an enhancement of a factor $4.6$.

\begin{center}%
\begin{tabular}
[c]{ll}%
{\includegraphics[
height=2.6541in,
width=4.0274in
]%
{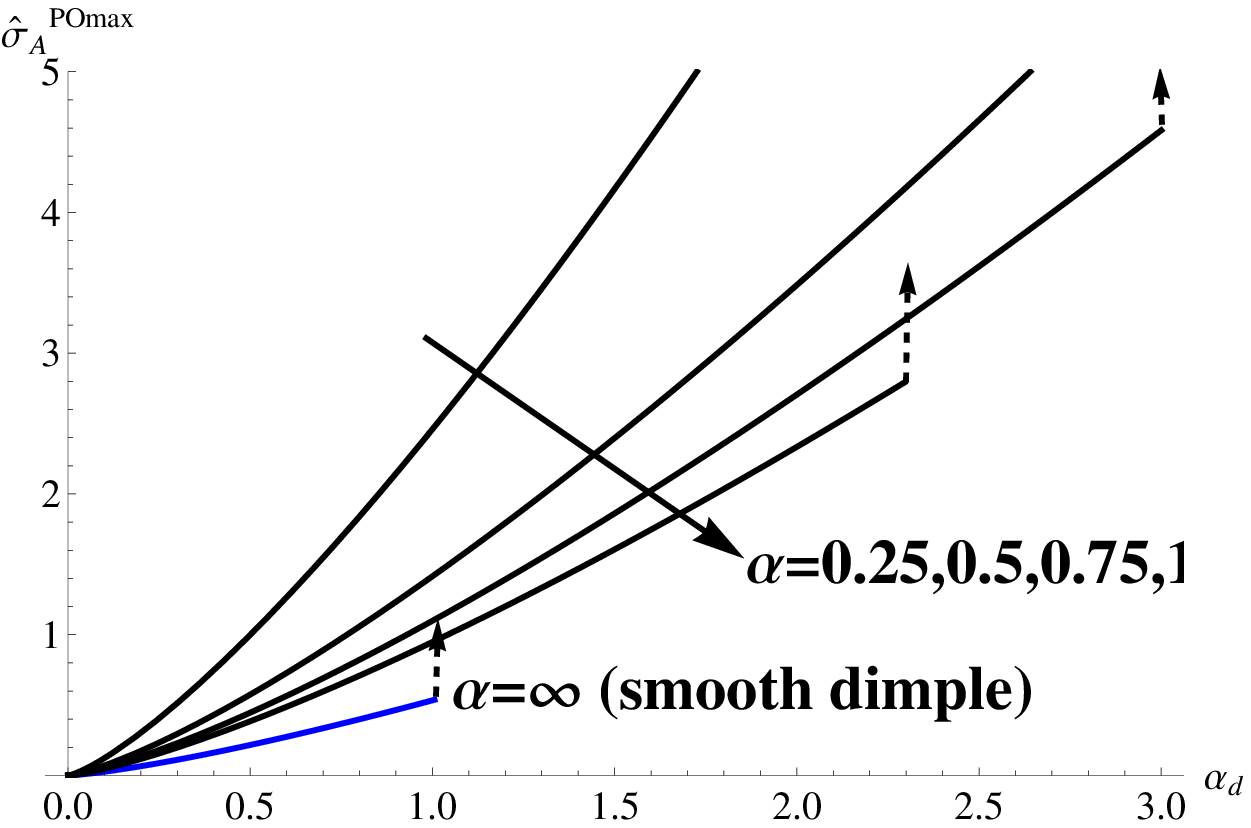}%
}
& (a)\\%
{\includegraphics[
height=2.5105in,
width=4.0274in
]%
{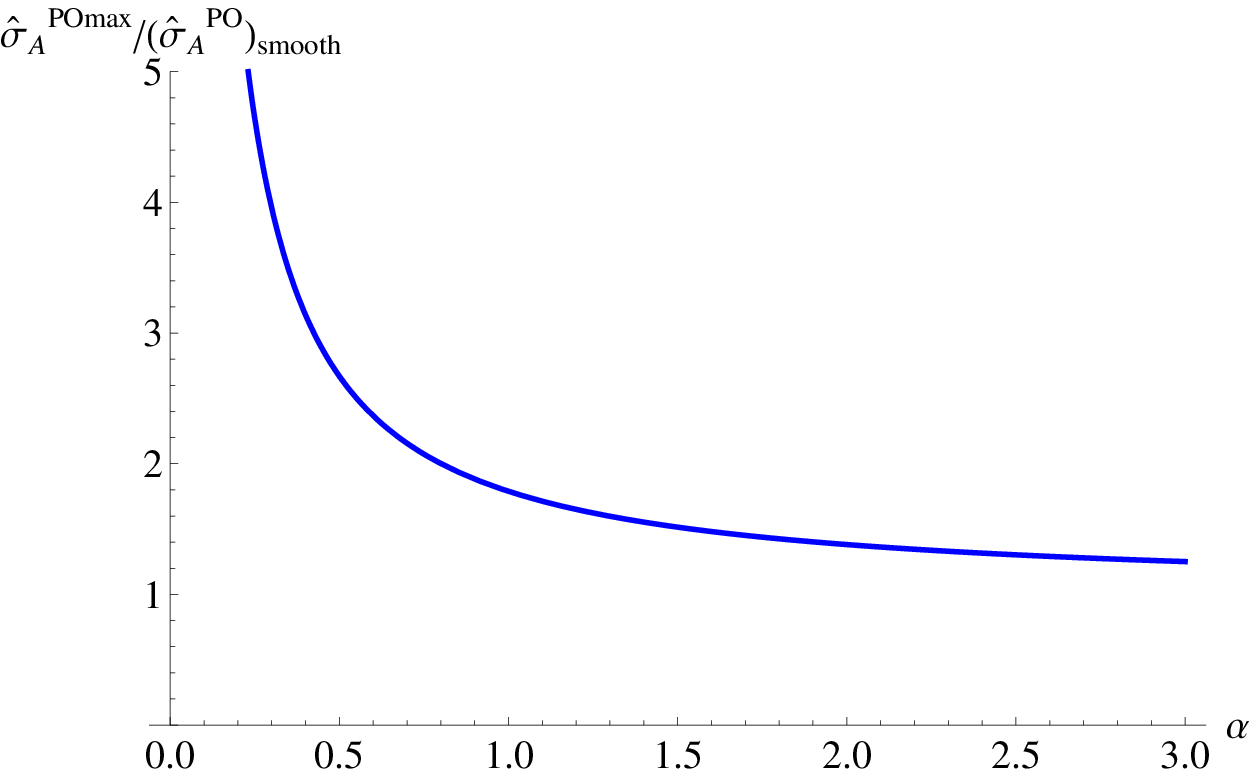}%
}
& (b)
\end{tabular}

Fig.6. (a) Pull-off $\widehat{\sigma}_{A}^{PO\max}$ for rough dimple as a
function of $\alpha_{d}$ for various $\alpha=0.75,1,\infty$ (smooth dimple).
(b) enhancement factor $\frac{\widehat{\sigma}_{A,rough}^{PO\max}}%
{\widehat{\sigma}_{A,smooth}^{PO}}$ as a function of $\alpha$
\end{center}

\subsection{Pressure-sensitive pull-off}

It is clear that, in the case of the smooth dimple, the equilibrium size of
the crack at zero load $\widehat{c}_{eq}$ is always smaller than the size at
pull-off, $\widehat{c}_{\max}$. Unfortunately, the expression for $\widehat
{c}_{eq}$ for the smooth dimple is a little lengthy%
\begin{equation}
\widehat{c}_{eq}=\frac{1}{12\alpha_{d}^{2}}\left[  \alpha_{d}^{4}+\frac
{\alpha_{d}^{8/3}\left(  24+\alpha_{d}^{4}\right)  }{p}+\alpha_{d}%
^{4/3}p\right]
\end{equation}
where $p=\left[  36\alpha_{d}^{4}+\alpha_{d}^{8}+24\left(  9+\sqrt{3}%
\sqrt{27+\alpha_{d}^{4}}\right)  \right]  ^{1/3}$. It can be approximated by
$\widehat{c}_{eq}=0.7/\alpha_{d}^{0.65}$ but only up to $\alpha_{d}<0.4.$

Now, for small roughness, i.e. large $\alpha$, the loading curve will be close
to the loading curve of the smooth case, and therefore the equilibrium crack
size $\widehat{c}_{eq}\left(  \alpha_{d,rough}^{loading}\right)  $ will be
increased only marginally,\ while the unloading curve will have moved the
point of pull-off to smaller sizes $\widehat{c}_{\max}\left(  \alpha
_{d,rough}^{unloading}\right)  .$ Using the power law approximations for
$\widehat{c}_{eq}\ $and $\widehat{c}_{\max}$, we obtain the coincidence and
pull-off is unique for
\begin{equation}
\alpha>\alpha_{uni}\simeq0.9
\end{equation}
A more exact calculation shows that $\alpha_{uni}$ varies by only few percent
depending on $\alpha_{d}$. Therefore, there is an interesting range below
$\alpha_{uni}$ which show pressure-sensitiveness. We shall now investigate a
little more this range.

For the pressure-sensitive pull-off range, $\alpha<\alpha_{uni}$, a simple but
good approximation (provided the wavelength of the roughness is short enough)
is that we load up to a certain $\widehat{\sigma}_{A}$, and since we end up in
the unloading unstable branch, we will have immediately pull-off. Hence,
loading up to
\begin{equation}
\widehat{\sigma}_{A,peak}=-1+\sqrt{1-\frac{1}{\widehat{c}_{peak}^{2}}}%
+\frac{\alpha_{d,rough}^{loading}}{\sqrt{\widehat{c}_{peak}}}%
\end{equation}
results in a pull off
\begin{equation}
\widehat{\sigma}_{A}^{PO}=-1+\sqrt{1-\frac{1}{\widehat{c}_{peak}^{2}}}%
+\frac{\alpha_{d,rough}^{unloading}}{\sqrt{\widehat{c}_{peak}}}%
\end{equation}

We need, in fact, to explore only the range
\begin{equation}
\widehat{c}_{\max}\left(  \alpha_{d,rough}^{unloading}\right)  <\widehat
{c}_{peak}<\widehat{c}_{eq}\left(  \alpha_{d,rough}^{loading}\right)
\end{equation}
as otherwise, the upper bound pull-off is obtained, which we have already estimated.

A set of results is shown in Fig.7. It is shown that the pull-off pressure
grows at first \textit{linearly with preload}, and then saturates to a value
which we have already estimated in (\ref{sigma-pull-off-max}) and Fig.6. If we
continue the curves above this value there would be a second transition
towards full contact, which however we did not include in the Figure, for
simplicity. In Fig7a,b,c, the case of $\alpha_{d}=0.5,1,1.5$ respectively are
represented, and an horizontal gray line indicates the pull-off value for the
smooth dimple. This permits to estimate the enhancement whose maximum value
corresponds to the scale in Fig.6. For larger values of $\alpha$, the maximum
enhancement is smaller, however the sensitivity to the pre-load is higher, so
it is possible to have higher pull-off, for a given preload.

\begin{center}%
\begin{tabular}
[c]{ll}%
{\includegraphics[
height=2.5105in,
width=4.0274in
]%
{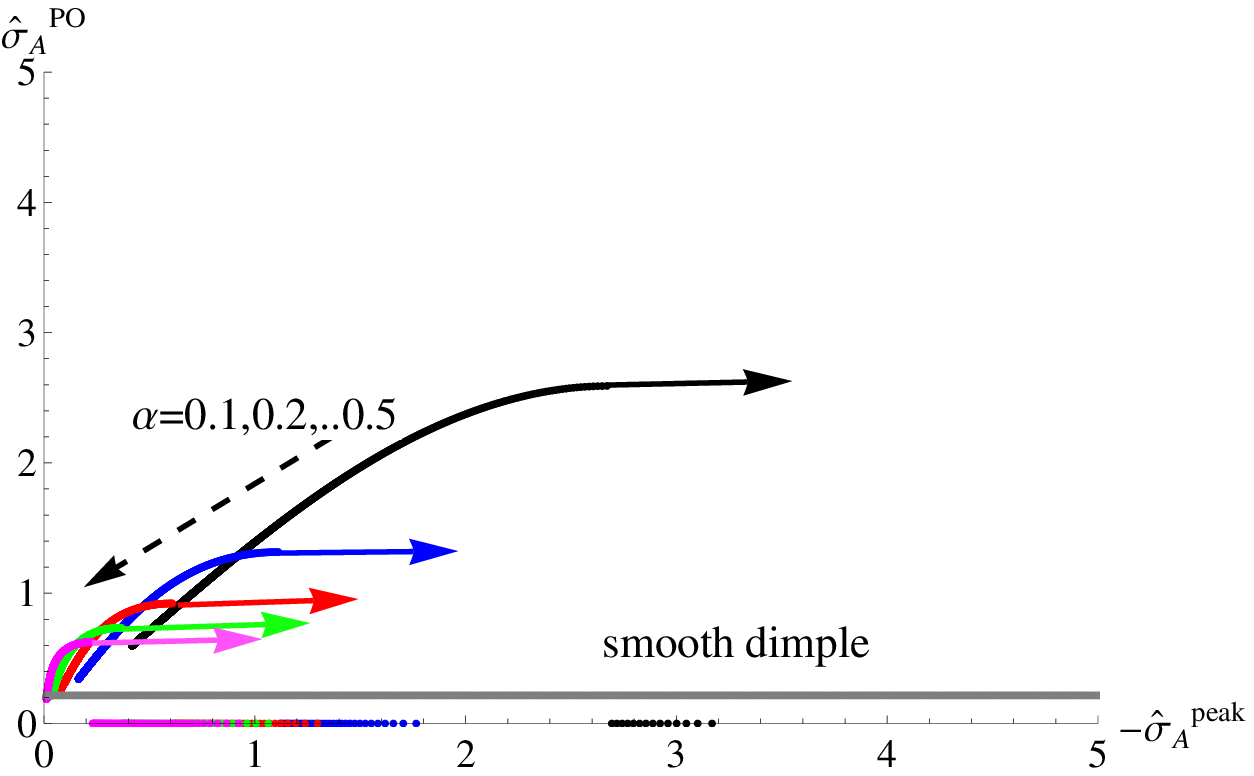}%
}
& (a)\\%
{\includegraphics[
height=2.4881in,
width=4.0274in
]%
{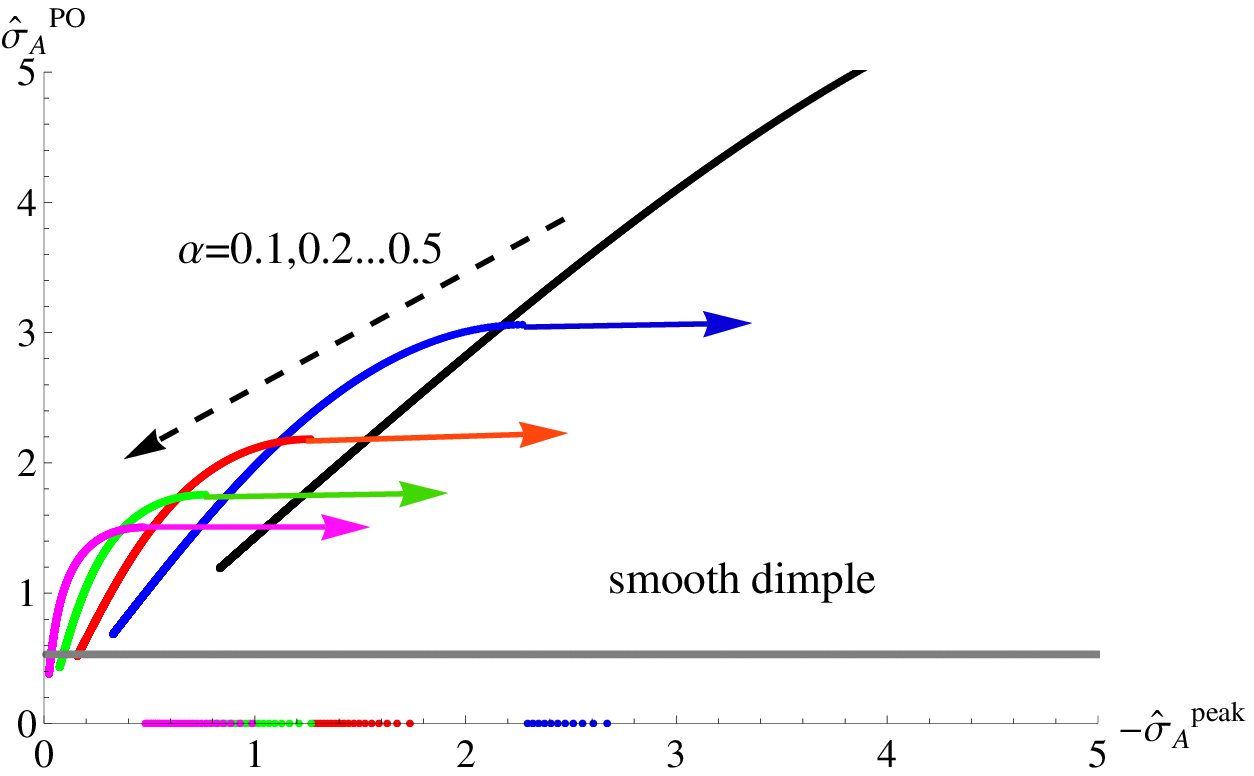}%
}
& (b)\\%
{\includegraphics[
height=2.4881in,
width=4.0274in
]%
{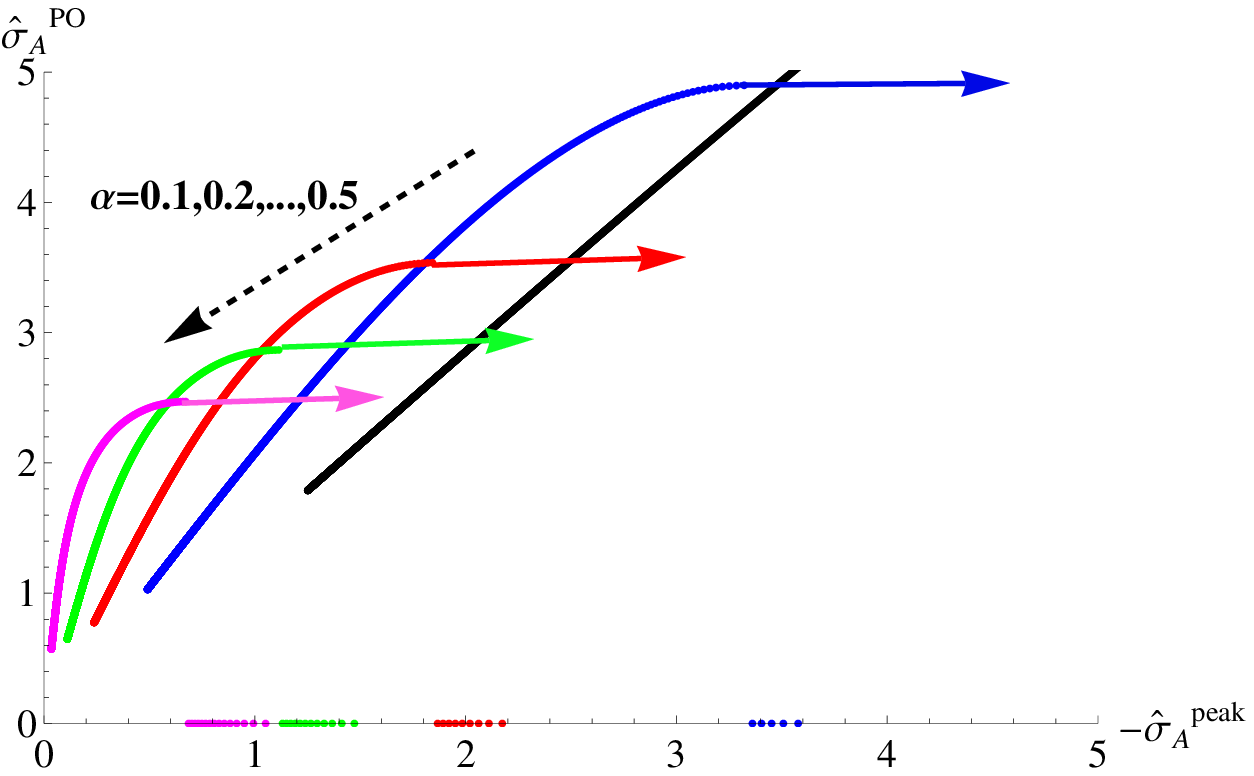}%
}
& (c)
\end{tabular}

Fig.7. Pressure-sensitive pull-off $\widehat{\sigma}_{A}^{PO}$ as a function
of preload $-\widehat{\sigma}_{A,peak}$ for three cases of $\alpha
_{d}=0.5,1,1.5$ (a,b,c, respectively)
\end{center}

\section{Discussion}

Regarding the validity of the solution, some of the conclusions reached in
Guduru (2007), but also Ciavarella (2016a, 2016b) translate into this problem,
and viceversa, the present solution clarifies some limitations of the Guduru
solution. In other words, we are assuming from the outset that a continuous
single connected (non-contact) area is obtained. When roughness is very large,
as is the case when $\alpha$ is low, we can imagine that this is not the way
the solution proceeds. Indeed, contact could be established only in the crests
and proceed from crest to crest, before a continuous contact is possible. This
explains why, \textit{counterintuitively}, we have reached the conclusion that
unbounded enhancement occurs for conditions of large roughness i.e. when
Johnson parameter is very low. Guduru's geometry (the sphere, in fact
approximated as a paraboloid) is such that one could postulate that any
enhancement should occur after sufficiently large pressure has been applied
--- although this at one point would involve finite strains in the sphere. In
Guduru's problem, $\alpha=0.57$ corresponds to an enhancement of the sphere
pull-off of a factor 4 --- see Ciavarella (2016a), whereas in our dimple case,
from (\ref{sigma-pull-off-max}) we have a value which is slightly smaller,
$2^{1.3}=\allowbreak2.\,\allowbreak46$ for $\alpha=0.57$. However, Guduru
\&\ Bull (2007) have actually demonstrated values of enhancement higher than
15 in experiments and there is no reason not to expect similar order of
magnitude also in our dimple case, justifying our figures above which contain
also the case $\alpha=0.1$. More precise estimates about this limit are not
obvious: as Guduru (2007) remarked the limit is not the monotonicity of the
profile. Another limit to the enhancement is that our analysis is limited to
the JKR regime of large soft materials with high adhesion.

We have not discussed in great details the detachment from full contact.
Johnson (1995) also does not discuss in detail for his sinusoidal waviness
case, referring to the fact that we need a tension of the order of theoretical
strength, and suggests air entrapment, contaminants, or indeed finer scale
roughness, may reduce this. In fact, even for a single scale of waviness, a
Maugis solution (Hui \textit{et al}, 2001, Jin \textit{et al}, 2016a, Jin
\textit{et al}, 2016b) shows that detachment will start when the \textit{peak
stress in the full contact state} reaches the theoretical strength, and in our
case this means
\begin{equation}
\widehat{\sigma}_{A}=\widehat{\sigma}_{0}-1-\widehat{p}^{\ast}%
\end{equation}
although the actual critical condition to open the contact will depend also on
the COD (Crack Opening Displacement) which has to reach the Maugis range of
attraction forces. However, this shows that when the sum of $T+p^{\ast}$ is
comparable to the theoretical strength, we can start opening already at values
much smaller than theoretical strength. This limit is not so remote, since it
occurs when the height of the dimple compared to its width (and/or the
amplitude of waviness compared to the wavelength), become of the order of 0.1.
Indeed, as theoretical strength is of the order of $\sigma_{0}\simeq
0.05E^{\ast}$, even for $g/\lambda=0.1$ we have $p^{\ast}=0.3E^{\ast}%
=6\sigma_{0}$.

In the presence of roughness and the shallow depression, the "strong adhesion"
regime due to full contact disappears in realistic cases, and it will depend
also on some appropriate Tabor parameter. Also the enhancement of the "weak
adhesion" regime will depend on appropriate Tabor parameter, but a full
solution to the problem requires a full numerical investigation, outside the
scopes of the present paper.

\section{Conclusions}

Originally, McMeeking \textit{et al.} (2010) introduced the "dimple" model as
a simple geometrical model to explain a bistable system realized with just
elastic materials and van der Waals adhesive forces, showing the possible
switch from "strong adhesion" realized when pushing in full contact from the
stable intermediate equilibrium, to "weak adhesion", when this pressure is not
impressed, and one has the pull-off from the partial contact state. The
analysis we conducted shows that, with roughness, the dimple model shows a
much higher resistance to jump into full contact, and therefore the "strong
adhesion" is obviously an ideal limit on two grounds: first, it may be
difficult to achieve due to geometrical imperfections, depressions, air
entrapment, contaminants, and roughness indeed; second, the separation from
this state may not be as difficult as expected, for the same very reasons.
However, we showed that, partially balancing this effect, we have an
"enhancement" of the "weak-adhesion" regime, which may serve the purpose if
one can calibrate the geometry. We have shown a reduced parametric dependence
of the system in the asymptotic expansion for small wavelength roughness on
only two dimensionless parameters, one being the Johnson parameter for the
sinusoid, and the other the corresponding Johnson parameter for the dimple. We
obtained that when roughness is relatively large, a pressure-sensitive region
is expected, and in this region, the actual pull-off depends monotonically on
the pre-load, and indeed in a significant regions of parameters, linearly on
pre-load. The model adds to our understanding of the effect of multiscale
roughness on adhesion, which remains a complex problem in the general case.

\section{Acknowledgements}

A.P. is thankful to the DFG (German Research Foundation) for funding the
project HO 3852/11-1.

\section{\bigskip References}

Ciavarella, M. (2016a), On roughness-induced adhesion enhancement, J. Strain
Analysis, in press, arXiv preprint arXiv:1602.06089.

Ciavarella, M. (2016b). An upper bound to multiscale roughness-induced
adhesion enhancement. Tribology International, 102, 99-102.

Fuller, K. N. G., \& Tabor, D. The effect of surface roughness on the adhesion
of elastic solids. Proc Roy Soc London A: 1975; 345:1642, 327-342

Gao, H., \& Yao, H. Shape insensitive optimal adhesion of nanoscale fibrillar
structures. Proceedings of the National Academy of Sciences of the United
States of America,2004: 101(21), 7851-7856.

Guduru, P.R. (2007). Detachment of a rigid solid from an elastic wavy surface:
theory J. Mech. Phys. Solids, 55, 473--488

Guduru, P.R. , Bull, C. (2007) Detachment of a rigid solid from an elastic
wavy surface: experiments J. Mech. Phys. Solids, 2007: 55, 473--488

Huber G, Gorb S, Hosoda N, Spolenak R, Arzt E. Influence of surface roughness
on gecko adhesion. Acta Biomater 2007;3:607--10.

Hui, C. Y., Lin, Y. Y., Baney, J. M., \& Kramer, E. J. (2001). The mechanics
of contact and adhesion of periodically rough surfaces. Journal of Polymer
Science Part B: Polymer Physics, 39(11), 1195-1214.

Hui, C. Y., Glassmaker, N. J., Tang, T., \& Jagota, A. (2004) Design of
biomimetic fibrillar interfaces: 2. Mechanics of enhanced adhesion. Journal of
The Royal Society Interface: 1(1), 35-48.

Jin, F., Guo, X., \& Wan, Q. (2016a). Revisiting the Maugis--Dugdale Adhesion
Model of Elastic Periodic Wavy Surfaces. Journal of Applied Mechanics, 83(10), 101007.

Jin, F., Wan, Q., \& Guo, X. (2016b). A double-Westergaard model for adhesive
contact of a wavy surface. International Journal of Solids and Structures,
102, 66-76.

Johnson KL, Kendall K., and Roberts A. D. (1971). Surface energy and the
contact of elastic solids. Proc Royal Soc London A: 324. 1558.

Johnson K.L. 1995. The adhesion of two elastic bodies with slightly wavy
surfaces, Int J Solids and Struct 32 (3-4), , pp. 423-430

Kamperman, M., Kroner, E., del Campo, A., McMeeking, R. M., \& Arzt, E.
Functional adhesive surfaces with \textquotedblleft gecko\textquotedblright%
\ effect: The concept of contact splitting. Advanced Engineering Materials,
2010: 12(5), 335-348.

Kesari, H., \& Lew, A. J. (2011). Effective macroscopic adhesive contact
behavior induced by small surface roughness. Journal of the Mechanics and
Physics of Solids, 59(12), 2488-2510.

Maugis, D (2000). Contact, adhesion and rupture of elastic solids (Vol. 130).
Springer, New York.

McMeeking, R. M., Ma, L., \& Arzt, E. (2010). Bi-Stable Adhesion of a Surface
with a Dimple. Advanced Engineering Materials, 12(5), 389-397.

Pugno NM, Lepore E. (2008) Observation of optimal gecko's adhesion on
nanorough surfaces. Biosystems;94:218--22.

\bigskip

\section{Appendix. On the use of plane strain approximation.}

In (\ref{plane-strain}), we used a plane strain approximation, under the
assumption that except perhaps for the first 2-3 oscillations, the problem
really is plane strain. However, here is a numerical proof. Obviously we meant
(\ref{plane-strain}) for a sinusoidal wave roughness of amplitude $g$ and
wavelength $\lambda$,%
\begin{equation}
f\left(  r\right)  =g\left(  1-\cos\left(  \frac{2\pi r}{\lambda}\right)
\right)
\end{equation}
If we take the standard cumulative superposition approach used by Guduru
(2007) and take the contact area $c\rightarrow\infty$, we get
\begin{equation}
h\left(  c\right)  =c\int_{0}^{c}\frac{f^{\prime}\left(  x\right)  dx}%
{\sqrt{a^{2}-x^{2}}}=\pi^{2}\frac{g}{\lambda}cH_{0}\left(  \frac{2\pi
c}{\lambda}\right)  \simeq-\pi\sqrt{\frac{c}{\lambda}}g\sin\left(  \frac{\pi
}{4}-\frac{2\pi c}{\lambda}\right)
\end{equation}
which we\ have expanded for small $\lambda$. Taking only the leading term in
the derivative $h^{\prime}\left(  x\right)  $ and using%
\begin{equation}
p\left(  r\right)  =\frac{E^{\ast}}{\pi}\int_{r}^{c}\frac{h^{\prime}\left(
x\right)  dx}{\sqrt{x^{2}-r^{2}}}%
\end{equation}
we find that
\begin{equation}
\frac{p\left(  r\right)  }{p^{\ast}}=2\sqrt{\frac{c}{\lambda}}\int_{r}%
^{c}\frac{\cos\left(  \frac{\pi}{4}-\frac{2\pi x}{\lambda}\right)  dx}%
{\sqrt{x^{2}-r^{2}}}%
\end{equation}

When we\ take $c=50\lambda$ (Fig.A1a) we\ find that the error is negligible
but perhaps still visible on the first oscillations, but this reduces further
with $c=100\lambda$ (see Fig.A1b), and as we are considering here the full
contact pressure when $c=\infty$, we do not need to worry about this approximation.

\begin{center}%
\begin{tabular}
[c]{ll}%
{\parbox[b]{5.0548in}{\begin{center}
\includegraphics[
height=3.064in,
width=5.0548in
]%
{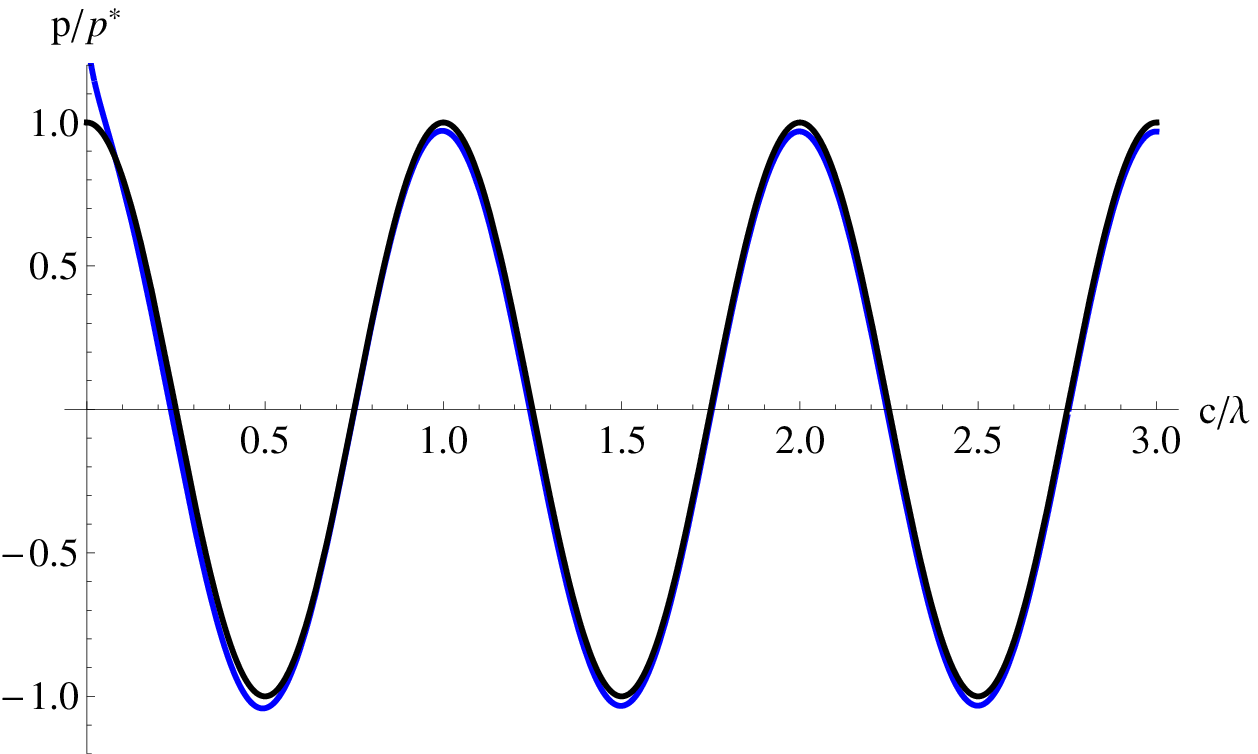}%
\\
{}%
\end{center}}}
& (a)\\%
\raisebox{-0pt}{\parbox[b]{5.0548in}{\begin{center}
\includegraphics[
height=3.064in,
width=5.0548in
]%
{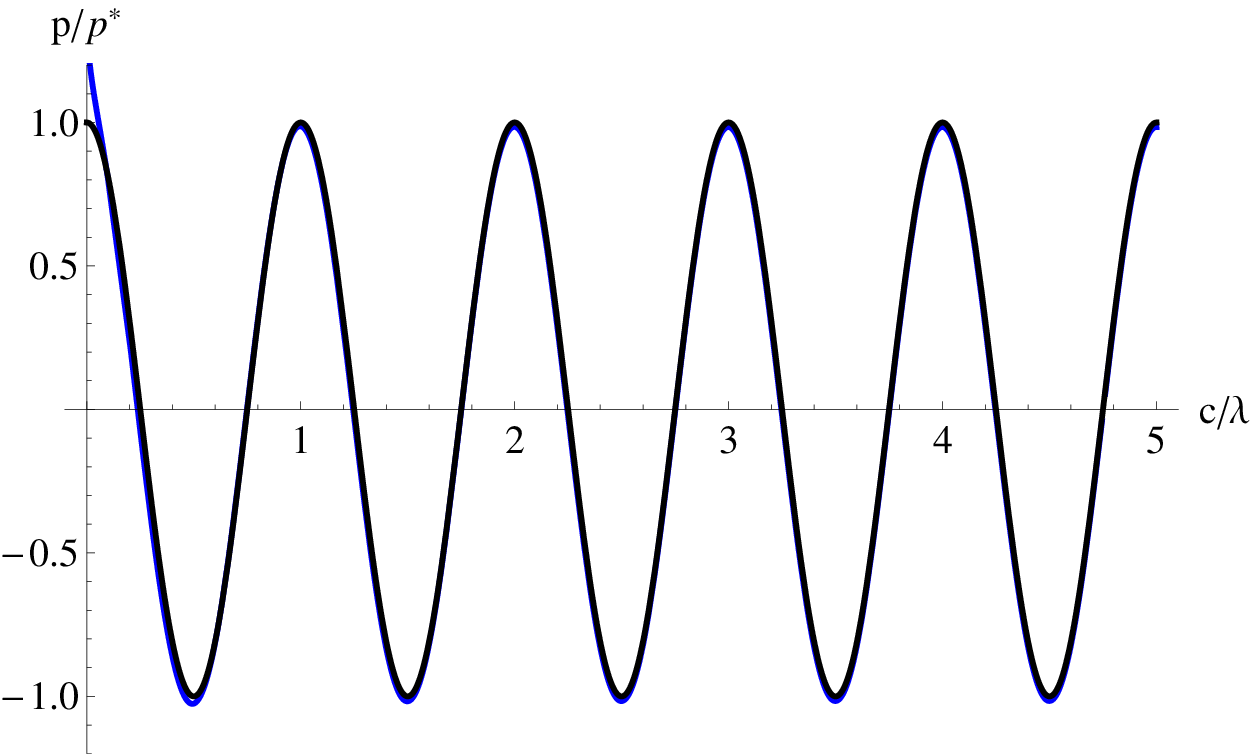}%
\\
As Fig.1, but for $c=100\lambda$.
\end{center}}}
& (b)
\end{tabular}

Fig. A1 Function $\frac{p\left(  r\right)  }{p^{\ast}}$ numerically obtained
(blue solid line) for $c=50\lambda$, against the plane strain solution
$\cos\left(  2\pi c/\lambda\right)  $. (solid line). (b) but for
$c=100\lambda$.
\end{center}

\end{document}